# Spectroscopic fingerprints of valence and spin states in manganese oxides and fluorides

*Ruimin Qiao[1,2], Timothy Chin [3], Stephen J. Harris[2], Shishen Yan[1], Wanli Yang[2]*

1. School of Physics, Shandong University, Jinan 250100, P. R. China
2. Advance Light Source, Lawrence Berkeley National Laboratory, Berkeley, CA 94720, USA
3. Environmental Energy Technologies Division, Lawrence Berkeley National Laboratory, Berkeley, CA 94720, USA

**Abstract**

We performed a systematic study of soft X-ray absorption spectroscopy in various manganese oxides and fluorides. Both Mn *L*-edges and ligand (O and F) *K*-edges are presented and compared with each other. Despite the distinct crystal structure and covalent/ionic nature in different systems, the Mn-*L* spectra fingerprint the Mn valence and spin states through spectral lineshape and energy position consistently and evidently. The clear O- and F- *K* pre-edge features in our high resolution spectra enable a quantitative definition of the molecular orbital diagram with different Mn valence. In addition, while the binding energy difference of the O-*1s* core electrons leads to a small shift of the O-*K* leading edges between trivalent and quadrivalent manganese oxides, a significant edge shift, with an order of magnitude larger in energy, was found between divalent and trivalent compounds, which is attributed to the spin exchange stabilization of half- filled *3d* system. This shift is much enhanced in the ionic fluoride system. This work provides the spectroscopic foundation for further studies of complicated Mn compounds.

## 1. Introduction:

Manganese compounds feature a variety of physical and chemical properties that are of both fundamental and practical interests. While the famous manganite compounds have been extensively studied for its colossal magnetoresistance[1], there has been increasing research interests on manganese compounds as low-cost and safe Lithium-ion battery positive electrodes recently[2, 3]. In particular, both spinel[2, 3] and Li-rich[4-6] manganese oxides have become new research focuses over the past several years. Mn- based battery electrodes hold the promise for high power and high capacity batteries dictated by the powering of electric vehicles, however, typically suffer cycling problems. Besides the issues with electrolyte, the lack of clear understanding of the high voltage (power) operation of Mn- based electrode, which involves the Mn and O electronic state evolution, has hindered the rational optimization of the materials. Therefore, a *direct* probe of valence states in these systems becomes critical for both fundamental understanding and practical advances of Mn compounds.

Previously, hard X-ray absorption of Mn *K*-edges has been extensively applied for studying the structure and valence of Mn in various compounds[6-9]. Although hard X-rays benefit from deep penetration depth and non-vacuum instrumentation, the dominating signal of Mn *K*-edge XAS is from the *4p* states, which are not the valence state of transition metals. The Mn-*3d* valence states only exhibit weak features through the low intensity dipole-forbidden quadrupole *s-d* transitions[6-9]. In principle, the electron energy-loss spectroscopy (EELS)[10] and hard x-ray Raman scattering[11] could provide the direct information of transition metal *3d* states, however, they both lack the resolution to reveal the detailed features in *3d* electronic structure. X-ray photoemission spectroscopy (XPS) is sensitive to the chemical potential, but for the insulating oxides and fluorides, conventional XPS is not reliable to define the small variation of about half eV [12] with Mn valence. On the other hand, soft X-ray absorption spectroscopy (XAS) probes directly the *3d* unoccupied valence states of Mn through the dipole *2p-3d* transitions[7, 13-15], as well as the ligand unoccupied states[14-16]. Indeed, we have recently demonstrated that high resolution soft X-ray XAS could directly probe the electronic structure of the *3d* states in $Li_xFePO_4$ cathode material[11].

In this paper, we provide a systematic study of high-resolution soft X-ray Mn *L*-edge and ligand (including O and F) *K*-edge XAS of Mn- oxides and fluorides. Previous studies of soft X-ray XAS on these compounds [13, 14, 16] discussed separately the Mn[13, 14] , O[14] , and F[16] edges based on surface-sensitive total electron yield (TEY) data. Our work presents the most complete and in-depth comparison between the systems among all the edges, and our experiments were performed with both TEY and bulk-sensitive total fluorescence yield (TFY) modes on samples with zero air-exposure through our special sample loading mechanics (Fig.1). Such comparison presents more information on the spectroscopic manifestation of manganese valence, spin states, and the covalent character of Mn-ligand bond. The data show the detailed electronic structure contrast in these systems and we deliver a quantitative molecular orbital diagram of the unoccupied valence states. Such high quality soft X-ray absorption spectra and analysis are aimed to build the spectroscopic foundation for further studying complex Mn compounds for practical applications[2-6].

## 2. Material and methods:

| compound | syngony | space group | Mn site symmetry | Mn formal valence | d-band occupancy |
|---|---|---|---|---|---|
| $MnO_2$ | tetragonal | $P4_2/mnm$ | distorted Octahedron | 4+ | $3d^3$ |
| $Mn_2O_3$ | cubic | 1a-3 | Octahedron | 3+ | $3d^4$ |
|  |  |  | distorted Octahedron |  |  |
| $MnF_3$ | monoclinic | C2/c | distorted Octahedron |  |  |
| MnO | cubic | fm-3m | Octahedron | 2+ | $3d^5$ |
| $MnF_2$ | tetragonal | $P4_2/mnm$ | distorted Octahedron |  |  |

Table 1: Basic information of manganese oxides and fluorides studied in this paper.

Soft X-ray XAS was performed at Beamline 8.0.1 of the Advanced Light Source (ALS) at Lawrence Berkeley National Laboratory (LBNL). The undulator and spherical grating monochromator supply a linearly polarized photon beam with resolving power up to 6000. Experiments were performed at room temperature and with the linear polarization of the incident beam 45◦ to sample surfaces. Data were collected in both surface-sensitive TEY and bulk-sensitive TFY mode. All the spectra have been normalized to the beam flux measured by a clean gold mesh. The experimental energy resolution is better than 0.15eV.

The manganese oxides and fluorides studied in this work are listed in table 1. Chemicals with the highest possible purity are purchased from Sigma-Aldrich. Samples were pressed onto conducting carbon tape and loaded into our special sample transfer chamber (Fig.1) inside high-purity argon glove box. The sample transfer chamber was then sealed and mounted directly onto our ultra-high vacuum chamber without any air-exposure to the sample surfaces. The carbon tape was completely covered by a thick layer of Mn compounds; no oxygen signal from the tape contributes to the oxygen signal reported here. Despite such anaerobic handing, we noticed surface oxidization effect for MnO sample on both Mn L-edge and O K-edge X-ray absorption spectra, indicating the commercially purchased MnO may have been oxidized already. Therefore, we measured the slightly oxidized surface of a polished Mn metal as standard MnO, as previously demonstrated by XPS experiments[17]. In the meantime, under the same anaerobic environment, the $MnF_2$ turns out to be much more stable and delivers reliable spectra of divalent manganese system.

## 3. Results and Discussion:

The Mn L-edge XAS directly probes the electron dipole transition from *2p* core level to the *3d* valence states. Thus, the spectra are sensitive to factors that change *3d* orbital splitting and occupation, such as spin configuration, ligand field and manganese valence[13]. Fig. 1(a) shows Mn L-edge XAS spectra for all manganese samples listed in table 1. Only TEY data is shown because TFY signals on Mn L-edge in manganese oxides samples are severely distorted by the O-K emission and do not represent the absorption cross-section on Mn L-edge[18]. The spectrum of MnO acquired from surface-oxidized Mn metal is consistent with the previous EELS results[10] but with much better resolution; while the surface of MnO sample is obviously oxidized to $Mn_2O_3$ as shown in Fig. 1(b). This confirms the previously observed Mn oxidation process as characterized by XPS technique[17].

Although our Mn-*L* data are generally consistent with previous reports[14], plotting the spectra of oxides and fluorides together delivers systematic information on valence, spin states, and covalence of the materials. Firstly, the spin-orbit interaction splitting *2p* core states into $2p_{1/2}$ and $2p_{3/2}$, which corresponds to the well separated $L_2$ and $L_3$ absorption features, so-called "white lines". It has been well established that the intensity ratio of $I(L_3)/I(L_2)$ is related to the *3d* metal spin states, which is defined by Hund's rule and modified by the crystal field, i.e., high spin states correspond to large branching ratio $I(L_3)/[I(L_3)+I(L_2)]$ [19]. The spectra in Fig.1 (a) clearly shows that the increasing branching ratio with the decreasing Mn formal valence, indicating the system evolve towards high spin states from $Mn^{4+}$ ($3d^3$) to $Mn^{2+}$ ($3d^5$). The overall large branching ratio compared with the statistical values for Mn also indicates all the samples have high-spin ground states[10]. In particular, because $Mn^{3+}$ is a Jahn-Teller scheme, the lattice distortion leads to a high-spin ground state mixed with low-spin configuration[20].

Secondly, despite of the distinct crystal structures of oxides and fluorides (table 1), the overall lineshape of fluorides resembles that of oxides with the same Mn valence, especially for MnO and $MnF_2$ (Fig.1a). The fluorides, however, show much sharper features in both $L_3$ and $L_2$ edges than oxides. This sharpness contrast stems from the more ionic character of fluorides, which yields more atomic like sharp features. The oxides, on the other hand, exhibit only broadened features, indicating their more covalent nature than fluorides.

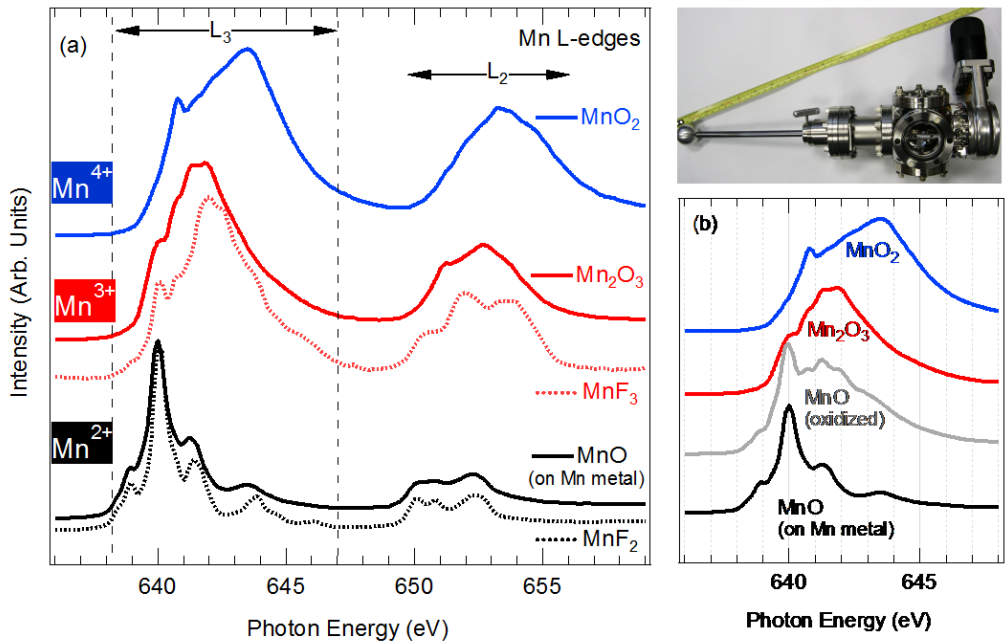

Fig. 1. (Color online) Photo shows the zero air-exposure sample transfer chamber that is ultra-high vacuum compatible. (a) Mn *L*-edge XAS spectra of all the Mn oxides and fluorides listed in table 1. The spectra were arranged according to the formal Mn valence. Oxides and fluorides were plotted in solid and dashed lines, respectively. (b) The Mn $L_3$-edge XAS spectra of all the Mn oxides.

Thirdly, in addition to the obvious lineshape changes, the gravity center of both the Mn $L_3$ and $L_2$ absorption features, as well as the leading edge position, exhibit systematic shift towards high energy with increasing Mn valence. Similar trend on MnO and $Mn_2O_3$ was observed before by *2p* core electron XPS spectra[21], and it is consistent with theoretical simulations[13, 22, 23]. Note that the oxides and fluorides are different on both the lattice structure and the covalent/ionic character; however, because XAS is only sensitive to the local environment of Mn rather than the periodic crystal lattice, both the lineshape and the energy position of the XAS consistently fingerprint the Mn valence in all the systems (Fig.1a). This consistency is crucial for studying practical materials with different structures, e.g., the spinel and layered Mn-based positive electrodes.

O *K*-edge XAS spectra of the Mn oxides collected with both TEY and TFY mode are displayed in Fig.2. In general, the spectra can be divided into two regions, the sharp "pre-edge" features below 534eV with the first peak marked as *a1* on TFY spectra, and the broad humps at higher energy. The pre-edge features originate from electron transitions between ligand *1s* core levels and partially filled manganese *3d* states, which are permitted due to the hybridization of O-*2p* and Mn-*3d* states[24]. The high energy humps are the broad O-*2p* conduction bands hybridized to Mn *4s,p* states.

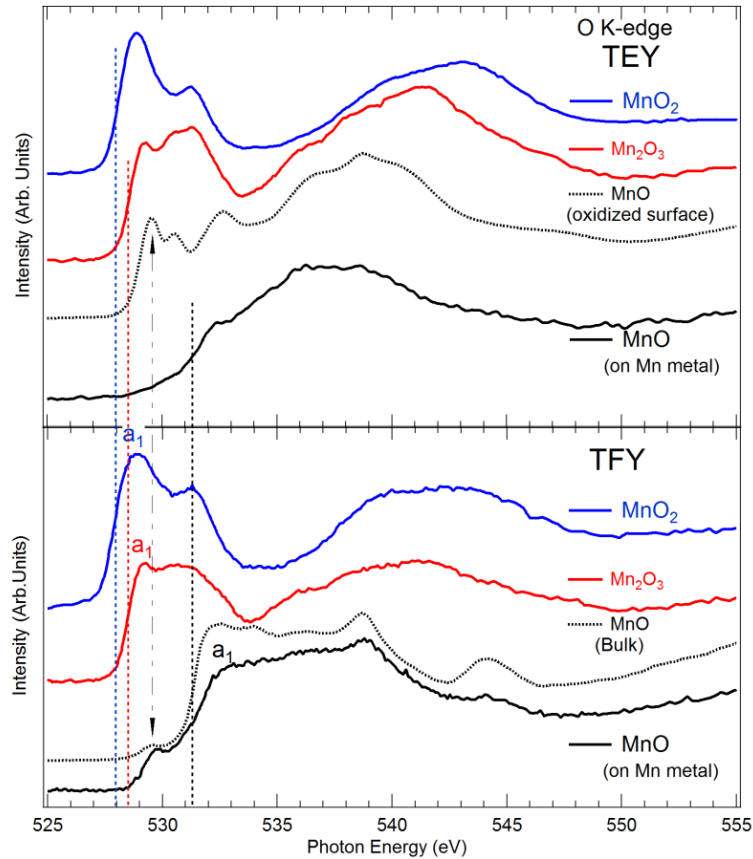

Fig. 2. (Color online) O *K*-edge XAS spectra of manganese oxides with different Mn oxidation states acquired by both TEY (top) and TFY (bottom) mode. Vertical dashed lines indicate the leading edge position of the spectra plotted in the same color.

For MnO, the surface-sensitive TEY spectrum of the powder is close to that of $Mn_2O_3$, while the bulk-probe TFY signal is similar to that of the oxidized Mn metal, both of which show weak features of $Mn_2O_3$ (black arrows in Fig.2). This suggests again that the oxidation product of MnO is $Mn_2O_3$, while the oxidation products of Mn metal is mainly MnO, consistent with the XPS findings[17]. We would like to mention that this is the first experimentally acquired bulk-sensitive O-*K* XAS spectra of MnO.

With the increasing Mn valence, the humps around 535-547eV shift towards higher energy. This trend is the same as that of the Mn-*L* XAS discussed above and could be explained by molecular orbital theory[14]. Interestingly, the absorption leading edge shows opposite shifts (Fig.2). $MnO_2$ exhibits the lowest leading edge, about 0.5eV lower than that of the $Mn_2O_3$. Because the O-*K* XAS leading edge corresponds to the lowest energy for exciting the O-*1s* electrons, this 0.5eV difference is consistent with the O-*1s* binding energy difference between $MnO_2$ (529.2eV) and $Mn_2O_3$ (529.7eV)[21]. However, the leading edge of MnO is 2.7eV higher than that of the $Mn_2O_3$, which is obviously too large to be compensated by the O-*1s* binding energy (529.3eV). As a matter of fact, such significant difference has been predicted by the Zaanen-Sawatzky-Allen (ZSA) diagram for half-filled $3d^5$ transition metals[25] and recently confirmed by XAS experiments[26]. For $3d^5$ systems, all the electron spins are aligned; to add (excite) an extra electron into $d^5$ states means it has to go in with opposite spin. This leads to a strong enhancement of both Coulomb U and charge transfer energies[25, 27], which pushes the absorption leading edge towards higher energy for several eV in $3d^5$ ($Mn^{2+}$) systems [26].

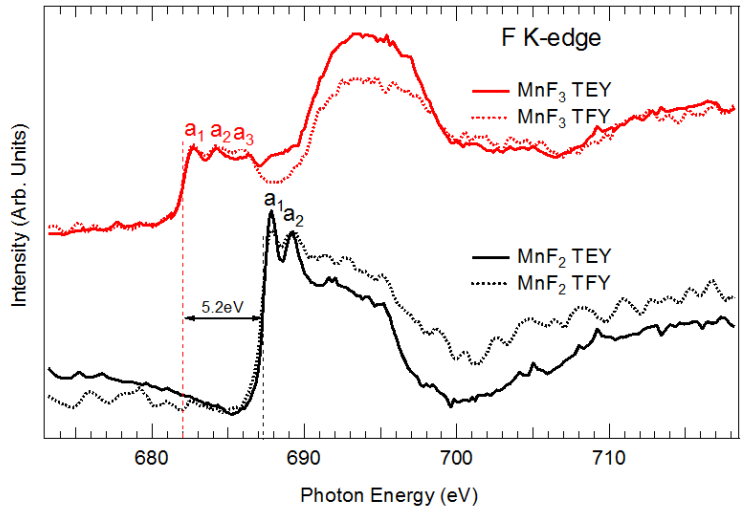

Fig. 3. (Color online) F *K*-edge XAS spectra of $MnF_3$ and $MnF_2$. The solid line represents TEY data, while the dash lines are TFY.

Such effect from exchange stabilization [25-27] is further enhanced in the more ionic fluoride systems. Fig.3 shows a 5.2eV energy shift of the leading edges between $MnF_3$ and $MnF_2$. The much localized electron states in ionic fluoride systems lead to much stronger Coulomb U penalty than that in the more covalent oxides. For the same reason, the features in F-*K* spectra are much sharper than that in O-*K* spectra, as also exhibited in the Mn-*L* spectra in Fig.1. Furthermore, the sharpness of the F-*K* pre-edge features allows us to clearly define three peaks, *a1*-*a3*, for $MnF_3$ and two peaks, *a1* and *a2*, for $MnF_2$ (Fig.3).

While the Mn-*L* spectra are dominated by atomic multiplet effects, the ligand-*K* pre-edge features correspond mainly to spin states and crystal fields[23]. This is because the pre-edge stems from the excitation from ligand *1s* core electron to the Legand-*2p*/Mn-*3d* hybridization states. As there is negligible overlap between ligand *1s* core hole and excited Mn-3*d* states, the

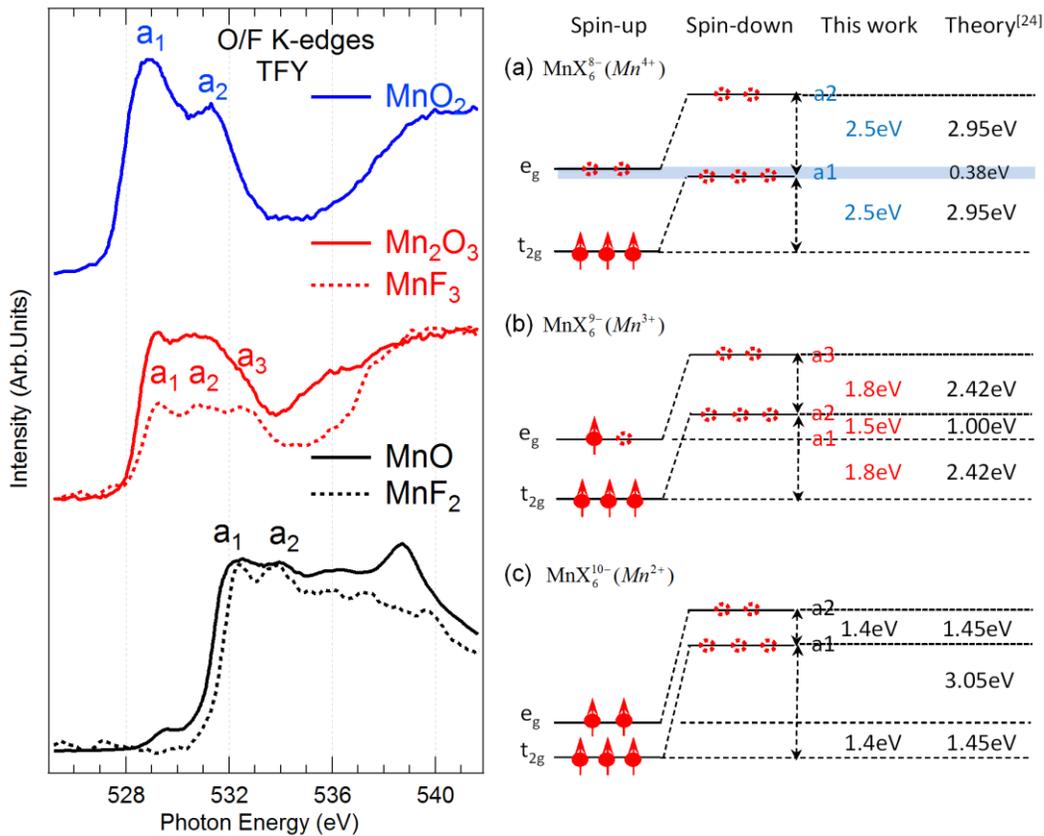

Fig. 4. (Color online). (left) Amplified O-*K* and F-*K* XAS pre-edge features of Mn oxides and fluorides. The energy of F-*K* spectra was arbitrarily shifted to align the *a1* feature with the *a1* of O-*K* spectra. (Right). Experimental molecular orbital diagrams based on the XAS spectra. The three clusters, $MnX_6^{8-}$, $MnX_6^{9-}$, and $MnX_6^{10-}$ (*X*=O, F, etc.) correspond to pre-edge features are not affected by atomic multiplet effects, and provide an opportunity to quantitatively define the molecular orbitals affected by crystal fields.

With the clearly defined multiple peaks in the pre-edge ligand-$K$ spectra, especially the F-$K$, we generated the quantitative molecular orbital diagram in Fig.4 by considering octahedral crystal field, $e_g$ and $t_{2g}$, splitting (Table 1). The overall diagram is based on the molecular orbital calculation by Sherman[28], so we also listed the calculated values in Fig.4 for comparison purpose. Although it seems the theory overestimates the crystal field effect for about half eV for $Mn^{4+}$ and $Mn^{3+}$, the spectroscopic results are qualitatively consistent with the theory: (1) the crystal field splitting between $t_{2g}$ and $e_g$ states increases with the Mn valence, while the exchange splitting between spin-up and spin-down states decreases. (2) For $Mn^{4+}$ ($3d^3$), although with three $3d$ unoccupied state levels, only two peaks are observable as a consequence of the nearly equivalent values of crystal field and exchange splitting. (3) For $Mn^{3+}$ ($3d^4$), triple-peak structure was observed from the transitions to $e_g$-up, $t_{2g}$-down, and $e_g$-down states. (4) As discussed above, $Mn^{2+}$ is a half-filled $3d^5$ system; the hybridization features are pushed towards higher energy and overlap with the broad conduction bands. Still, the relatively small splitting of the two clear features in $MnF_2$ agrees with the small value in theory, which is coupled with the huge exchange energy due to the exchange stabilization effect[25, 26].

The comparison of the overall ligand-$K$ spectral lineshape again indicates the difference on the ionic and/or covalent nature of the systems. $MnF_3$ exhibits relatively weak hybridization features since it is more ionic than $Mn_2O_3$. Interestingly, the increasing intensity of the hybridization pre-edge features in $MnO_2$ than $Mn_2O_3$ suggests the increasing covalent character with increasing Mn valence. This change naturally explains the evolution of the crystal field effect, which enhances with Mn valence as evidenced by the energy difference of the pre-edge features and the derived orbital diagram (Fig.4).

## 4. Conclusions:

This work provides a comprehensive comparison of Mn-$L$, O-$K$ and F-$K$ XAS spectra in various manganese oxides and fluorides with the formal Mn oxidation states, 4+, 3+ and 2+. Mn-$L$ spectra fingerprint the Mn valence despite of the very different crystal structure and ionic/covalent nature in these systems. We show that the spin states of Mn-3d electrons manifest themselves in both the branching ratio of Mn-$L$ spectra and the pre-edge splitting of ligand-K spectra. The comparison of the spin exchange stabilization in half-filled $3d^5$ systems, MnO and $MnF_2$, is also discussed.

Fluorides exhibit much sharper features in spectroscopy due to their ironic nature. More interestingly, the covalent character of the Mn-ligand bond was shown to increase with the Mn valence, leading to much enhanced crystal field effect on the Mn-$3d$ states.

Eventually, we have derived a quantitative molecular orbital diagram based on the spectroscopic data. The diagram is qualitatively consistent with theoretical calculations with quantitative discrepancy on the crystal field effect. Our results show that soft X-ray absorption spectroscopy is sensitive to the local environment, and could reveal abundant information on the key electronic structures that are relevant to practical properties, e.g., valence and spin states. It is thus a powerful tool for studying complicated Mn based materials with different crystal structures, e.g., spinel and layered Mn based positive electrodes for Lithium batteries.

**Acknowledgement:** This work is partially supported by the LDRD program at the LBNL. Works in China are supported by National Science Foundation No.51125004 and No.10974120. The Advanced Light Source at Lawrence Berkeley National Laboratory was supported by the U. S. Department of Energy under Contract No. DE-AC02-05CH11231.